\documentclass{elsart}

\usepackage{natbib}

\usepackage{amssymb}

\begin{document}

\begin{frontmatter}



\title{The $\Lambda \Lambda \to YN$ Weak Transition in Double-$\Lambda$ Hypernuclei}


\author[label1]{K.~Sasaki},
\author[label2]{T.~Inoue}, and
\author[label3]{M.~Oka}

\address[label1]{Institute of Particle and Nuclear Studies, KEK, 
                 1-1, Oho, Tsukuba, 305-0801 Ibaraki, Japan}
\address[label2]{Institut f\"ur Theoretische Physik, Universit\"at T\"ubingen
                 Auf der Mogenstelle 14, D-72076 T\"ubingen, Germany}
\address[label3]{Department of physics, Tokyo Institute of Technology,
                 2-12-1, Ookayama, Meguro-ku, 152-8551 Tokyo, Japan}

\begin{abstract}
We calculate the $\Lambda \Lambda \to YN$ transition rate of 
 ${^{\phantom{\Lambda}6}_{\Lambda \Lambda}}$He
 by the hybrid picture, the $\pi$ and $K$ exchanges plus the direct quark processes.
It is found that the hyperon-induced decay is weaker than the nucleon-induced decay,
 but the former may reveal the short-range mechanism of the weak transition
 and also give a clear signal of the strong $\Delta I=3/2$ transition.
The $\Lambda \Lambda \to Y \! N$ transition in double-$\Lambda$ hypernucleus 
 is complement to the $\Lambda N \to NN$ transition as it occurs 
 only in the $J=0$ channel, while the $J=1$ transition is dominant 
 in the $\Lambda N \to NN$ case.
\end{abstract}

\begin{keyword}
Hypernuclei \sep Nonmesonic decay \sep Direct quark process
\PACS 21.80.+a \sep 12.39.-x
\end{keyword}

\end{frontmatter}


\section{Introduction}
Recent analyses of hypernuclear nonmesonic decay reveal the importance of short-range 
 interaction of baryonic weak transition 
 \cite{ITO:NPA,Par:PRC,IOMI:NPA,SIO01:NPA,Jid:NPA,Ito:PRC,SIO02:NPA}. 
It is, in fact, realized that the momentum transfer is large 
 due to the $\Lambda-N$ mass difference, and therefore 
 the short range interaction becomes important.
We proposed to treat the short range part of $\Lambda N \to NN$ interaction using the valence 
 quark picture of the baryon and the effective four-quark weak Hamiltonian 
 \cite{ITO:NPA,IOMI:NPA,SIO01:NPA,SIO02:NPA}.  
We found that the direct quark (DQ) process gives significantly large contribution 
 and shows qualitatively different features from the meson exchange contribution, 
 especially in its isospin structure.
The kaon exchange contribution also largely enhance the $\Gamma_{nn}/\Gamma_{pn}$ ratio
 due to its spin structure
 \cite{Par:PRC,SIO01:NPA,Jid:NPA,SIO02:NPA}.\par

There remains, however, another problem, i.e.,
 experimental values of the proton asymmetry from polarized hypernucleus are incompatible 
 with the theoretical values, which is closely related to the $J=0$ transition amplitudes \cite{Nab:PRC}.
The $J=0$ amplitude is relatively small and therefore masked by the $J=1$ amplitude,
 but it is important to clarify the breaking of the $\Delta I=1/2$ rule 
 for the nonmesonic weak decay of hypernuclei.
In fact, 
 the quark model treatments of $\Lambda N \to NN$ transition predict large breaking
 of the $\Delta I =1/2$ rule in $J=0$ transition channel \cite{ITO:NPA,IOMI:NPA,SIO02:NPA,Mal:PhL}.

For the case of double-$\Lambda$ hypernuclei, 
 the nonmesonic decay is induced not only by the nucleon but also by the other $\Lambda$ hyperon:
 $\Lambda N \to NN ~ ~ {\rm{and}} ~ ~ \Lambda \Lambda \to YN$.
The two-body part of the nonmesonic decay rate of double-$\Lambda$ hypernuclei is classified as 
\begin{eqnarray}
\Gamma_{nm} 
 &=& 
   \Gamma_{NN} \hspace*{21.4mm} + \Gamma_{YN} \nonumber \\
 &=& 
   \Gamma_{nn} + \Gamma_{pn} (+ \Gamma_{pp}) + \Gamma_{\Lambda n} + \Gamma_{\Sigma^- p} + \Gamma_{\Sigma^0 n}
\end{eqnarray}
where $\Gamma_{pp}$ stands for the $\Sigma^+ p \to pp$ transition which is available 
 when the virtual $\Sigma$  exists \cite{SIO02:NPA}.
The $\Gamma_{YN}$ represents the $\Lambda \Lambda \to YN$ transition, which
 is a characteristic transition for the multi-$\Lambda$ system.
Experimentally, the $\Lambda \Lambda \to YN$ transition
 should be distinguishable from the $\Lambda N \to NN$ 
 decay mode because the hyperon with high momentum is emitted once it occurs 
 in double-$\Lambda$ hypernucleus.
Thus the hyperon-induced decay mode 
 provides us with valuable information on the $|\Delta S|=1$ baryon-baryon interaction 
 \cite{IUM:NPA,PRB:PRC}.

In sec.\ref{SEC:TrR},
 the hyperon-induced decay rate, $\Gamma_{YN}$, is expressed in terms of two-body amplitudes
 corresponding to the $\Lambda \Lambda \to YN$ transition.
In sec.\ref{SEC:Pot},
 we construct the meson exchange and DQ potential for the $\Lambda \Lambda \to YN$ transition.
In sec.\ref{SEC:Res},
 we show the result of the $\Lambda \Lambda \to YN$ transition rate 
 in ${^{\phantom{\Lambda}6}_{\Lambda \Lambda}}$He.
Finally, sec.\ref{SEC:Con} is devoted for the conclusion of this work.

\section{The $\Lambda \Lambda \to Y \! N$ transition rate \label{SEC:TrR}}
For the nonmesonic weak decay of double-$\Lambda$ hypernuclei,
 the hyperon induced decay mode, $\Lambda \Lambda \to YN$, 
 becomes possible besides the usual $\Lambda N \to NN$ transition.
Now we concentrate on the novel $\Lambda \Lambda \to YN$ decay.
The $\Lambda \Lambda \to Y \! N$ transition rate is given by
\begin{equation}
\Gamma_{Y \! N} = 
 \int \! \! \frac{d^3{p_1'}}{(2 \pi)^3} 
 \int \! \! \frac{d^3{p_2'}}{(2 \pi)^3}
 \frac{1}{2J_H+1} 
 \sum_{all} (2 \pi) \delta(E_f - E_i)
 | M_{fi} |^2
\label{EQ:decayrNR}
\end{equation}
where $M_{fi}$ is the $\Lambda \Lambda \to Y \! N$ transition amplitude, 
 $J_H$ is the total spin of initial hypernucleus, and 
 ${p'}_1$ and ${p'}_2$ are momenta of emitted particles, {\it{i.e.}}, hyperon and nucleon.
The summation indicates a sum over all quantum numbers of the initial and final particle systems.
The $\delta$-function for the energy conservation is given by
\begin{equation}
\delta(E_f \! - \! E_i) =
 \delta \left(
          M_Y + M_N + \frac{{p_1'}^2}{2M_Y} + \frac{{p_2'}^2}{2M_N}
         -2M_\Lambda -\epsilon_{\Lambda \Lambda}
        \right)
\end{equation}
where the hyperon binding energy is taken as $\epsilon_{\Lambda \Lambda} \simeq 7.25$MeV
 \cite{NAG:PRL}.
After the decomposition of angular momentum, the explicit form of $M_{fi}$ is 
\begin{equation}
|M_{fi}|^2 =
 (4 \pi)^4 \left|
 \int \! \! \! \! \int \! \! \! \! \int
 \Psi^{L'S'J}_f(R, r')
 {V}^{L L'}_{S S'J}(r,r')
 \Psi^{LSJ}_i(R, r)
 r^2dr {r'^2}dr' R^2dR
 \right|^2
\label{EQ:nonrelM}
\end{equation}
where ${V}^{L L'}_{S S'J}(r,r')$
 is the transition potential and
 $\Psi^{LSJ}(R, r)$ 
 is the wave function of the $\Lambda \Lambda$ or YN two-body system in the configuration space.
The $L$, $S$, and $J$ indicate the angular momentum, spin, and total spin for two-body system, respectively.

In order to extract the $\Lambda \Lambda$ state from the double-$\Lambda$
 hypernucleus,
 we assume, for simplicity, that the $\Lambda \Lambda$ state with $J=0$
 and $I=0$ is decoupled from the residual nuclei:
\begin{equation}
|{^{ \ A}_{\Lambda \Lambda}}Z \rangle = 
 | (\Lambda \Lambda)^{I=0}_{J=0} \rangle \otimes |{^{A-2}}Z \rangle 
\end{equation}
where the double-$\Lambda$ state is always in ${^1}S_0$ state 
 in the ground state.
The important point of the study the $\Lambda \Lambda \to YN$ process is that 
 the initial $\Lambda \Lambda$ pair is in the ${^1}S_0$ state and thus
 the $\Lambda \Lambda \to YN$ transition occurs only in the $J=0$ channel.
The $J=0$ transition amplitude is not well studied in the $\Lambda N \to NN$ decay,
 because the $J=1$ amplitude is much larger.
In contrast, the $\Lambda \Lambda \to YN$ decay directly probes the $J=0$ transition amplitude
 and therefore plays a complementary role.

The initial double-$\Lambda$ state is written in terms
 of the shell model wave function of $0s$ state :
\begin{equation}
\Psi_{\Lambda \Lambda}^{000}({\vec{r}_{\Lambda_1}},{\vec{r}_{\Lambda_2}}) = 
    \left[ \phi_{0s}^b({\vec{r}_{\Lambda_1}}) 
    \times
    \phi_{0s}^b({\vec{r}_{\Lambda_2}}) \right]^{J=0}
  = \left[ \phi_{0s}^{b_r}(\vec{r}) 
    \times
    \phi_{0s}^{b_R}(\vec{R}) \right]^{J=0}
\label{EQ:wf}
\end{equation}
where $\phi_{0s}^b$ is assumed to be a Gaussian with size parameter $b$.
The last expression of eq.(\ref{EQ:wf}) is the wave function in terms of 
 the relative and center-of-mass coordinate
 with $\vec{r} = \vec{r_{\Lambda_1}} - \vec{r_{\Lambda_2}}$ and 
 $\vec{R} = (\vec{r_{\Lambda_1}} + \vec{r_{\Lambda_2}})/2$.
The $b_r$ and $b_R$ are $\sqrt{2} b$ and $b / \sqrt{2}$, respectively.
The wave function for the final $YN$ state is assumed as the plane wave
 with $\vec{K}=\vec{r}_{Y} + \vec{r}_{N}$ and 
 $\vec{k}= (m_N {\vec{p}'}_{1} - m_Y {\vec{p}'}_{2})/(m_Y + m_N)$.
The $b$ parameter is taken as $1.9$fm which fit on the r.m.s. radius of $\Lambda$
 in ${^{\phantom{\Lambda}6}_{\Lambda \Lambda}}$He
 \cite{AN:Pcom}.

In addition, we introduce phenomenological correlation functions.
For the initial $\Lambda \Lambda$ wave function,
 we multiply to the uncorrelated wave function 
\begin{equation}
f_{\Lambda \Lambda}(r) = 
 \left( 1-e^{-\frac{r^2}{a^2}} \right)^n + br^2 e^{-\frac{r^2}{c^2}}
\label{EQ:SRCLN}  
\end{equation}
with $a=0.8$fm, $b=0.12$fm$^{-2}$, $c=1.28$fm, $n=1$, according to ref.\cite{PRB:PRC}.
On the other hand, 
 for the final $Y \! N$ state, we multiply
\begin{equation}
f_{YN}(r) = 1-j_0( q_c r ) 
\end{equation}
where $j_0(x)$ is a spherical Bessel function with $q_c=3.93$ fm$^{-1}$.

\section{Transition potential \label{SEC:Pot}}
In order to describe the weak $\Lambda \Lambda \rightarrow YN$ transition,
 we adopt the quark-meson hybrid model, which is focused on the quark degrees of freedom
 in a short-range region.
At long and medium distances, the transition is induced mainly by
 the one pion exchange (OPE) and one kaon exchange (OKE)
 processes.
The general form of $\Lambda \Lambda \rightarrow YN$ transition potential 
 in the meson exchange picture
 is given by
\begin{equation}
 V_{\Lambda \Lambda \to YN}(\vec q)
  =
  g_s [\bar{u}_Y \gamma_5 u_\Lambda]
    {\frac{1}{{\vec q}^2 + {\tilde m}_{i}^2}}
    \left( {  \frac{ \Lambda_{i}^2 -{\tilde m}_{i}^2 }
              { \Lambda_{i}^2 + {\vec q}^2   }
     } \right)^2
  g_w
    [\bar{u}_N (A+B \gamma_5) u_\Lambda]
  \label{eq:OPE}
\end{equation}
 where the coupling constants $g_s$, $g_w$, $A$ and $B$, 
 shown in Table \ref{TAB:SWcc},
 are chosen properly for each transition.
A monopole form factor with cutoff parameters, 
 $\Lambda_{\pi}$ = 800 MeV and $\Lambda_{K} =1300$ MeV, 
 are employed for each vertex.
As the energy transfer is significantly large, 
 we introduce the effective meson mass:
\begin{equation}
 \tilde{m} = \sqrt{m^2 - (q^0)^2}, 
 \hspace*{5mm}
 q^0 = 
 \left\{ 
 \begin{array}{rl}
 88.5 {\rm{MeV}} &
 {\rm{for}} \ \Lambda \Lambda \to \Lambda n \ {\rm{transition}} \\
 50.2 {\rm{MeV}} &
 {\rm{for}} \ \Lambda \Lambda \to \Sigma N \ {\rm{transition}}
 \end{array}
 \right. .
\end{equation}
The kaon exchange potential can be constructed similarly.
Both the strong and weak coupling constants are evaluated under the
 assumption of the flavor SU(3) symmetry.
Note that, for OKE, it involves the strangeness transfer and thus
 the strong and weak vertices should be exchanged.
\begin{table}
\caption{The strong and weak coupling constants for the $\pi$ and $K$ exchange potentials.
         The strong coupling constants are taken from Nijmegen strong couplings
         \cite{NSC97:PRC}.
         The weak coupling constants are in unit of 
         $g_w \equiv G_F m_\pi^2 = 2.21 \times 10^{-7}$.}
\begin{center}
\begin{tabular}{|c||llr|llr|llr|}
\hline
\hline
Meson & \multicolumn{3}{c|}{Strong c.c.} & \multicolumn{6}{c|}{Weak c.c.} \\ 
\cline{5-10}
 & & & & \multicolumn{3}{c|}{PC} & \multicolumn{3}{c|}{PV} \\ 
\hline
$\pi$ 
 & $g_{\Lambda \Sigma \pi}$   & = & $ 12.04$ 
 & $A_{\Lambda n \pi^0}$      & = & $ -1.05$ 
 & $B_{\Lambda n \pi^0}$      & = & $  7.15$ \\
 & & &  
 & $A_{\Lambda p \pi^-}$      & = & $  1.48$ 
 & $B_{\Lambda p \pi^-}$      & = & $-10.11$ \\
\hline
$K$   
 & $g_{\Lambda NK}$           & = & $-17.65$ 
 & $A_{\Lambda \Lambda K^0}$  & = & $  0.67$ 
 & $B_{\Lambda \Lambda K^0}$  & = & $ 12.72$ \\
 & & &  
 & $A_{\Lambda \Sigma^- K^+}$ & = & $ -0.55$ 
 & $B_{\Lambda \Sigma^- K^+}$ & = & $ -8.42$ \\
 & & &  
 & $A_{\Lambda \Sigma^0 K^0}$ & = & $  0.39$ 
 & $B_{\Lambda \Sigma^0 K^0}$ & = & $  5.95$ \\
\hline
\hline
\end{tabular}
\end{center}
\label{TAB:SWcc}
\end{table}

For shorter distances, we employ the direct quark (DQ) transition 
potential based on the constituent quark picture of baryons.  In this 
mechanism, a strangeness changing weak interaction between two 
constituent quarks induces the transition of two baryons.

The DQ potential is given as a nonlocal form as
\begin{equation}
   {V_{DQ}}^{L L'}_{S S'J}(r,r')
    =  -{G_{F}\over\sqrt{2}} \times W \, \sum_{i=1}^{7}
    \left\{ V_{i}^f f_{i}(r, r') + V_{i}^g g_{i}(r, r') 
    + V_{i}^h h_{i}(r, r') 
    \right\}
\end{equation}
where $r$ ($r'$) stands for the radial part of the relative coordinate 
 in the initial (final) state.
The explicit forms of $f_i$, $g_i$, and $h_i$ are given in ref. \cite{SIO01:NPA},
 and the coefficients, $V^k_{i}$, for the $\Lambda \Lambda \to YN$ transitions are 
 listed in Table \ref{TAB:Coef}.

\begin{table}
\caption{The coefficients, $V^k_{i}$, for $\Lambda \Lambda \to Y \! N$ transition.}
\begin{center}
\begin{tabular}{p{0.3cm}p{0.1cm}p{0.3cm}c|rrrrrrr}
\hline
\hline
 \multicolumn{3}{c}{$\Lambda \Lambda \to \Lambda n$} & $k$ 
 & \multicolumn{1}{c}{$V^k_{1}$} & \multicolumn{1}{c}{$V^k_{2}$} 
 & \multicolumn{1}{c}{$V^k_{3}$} & \multicolumn{1}{c}{$V^k_{4}$} 
 & \multicolumn{1}{c}{$V^k_{5}$} & \multicolumn{1}{c}{$V^k_{6}$}
 & \multicolumn{1}{c}{$V^k_{7}$} \\
\hline
 ${^1S_0}$ & $\to$ & ${^1S_0}$ & $f$
 & -0.3142 &  0.0595 & -0.0298 &  0.0003 & -0.0298 & -0.0336 & -0.0568 \\
           & $\to$ & ${^3P_0}$ & $g$
 & -0.0002 & -0.0002 &  0.0583 &  0.0212 &  0.0046 &  0.0050 &  0.0140 \\
           &       &           & $h$
 & -0.0002 & -0.0002 & -0.0193 & -0.0002 & -0.0143 &  0.0236 &  0.0192 \\
\hline
\hline
 \multicolumn{3}{c}{$\Lambda \Lambda \to \Sigma^0 n$} & $k$ 
 & \multicolumn{1}{c}{$V^k_{1}$} & \multicolumn{1}{c}{$V^k_{2}$} 
 & \multicolumn{1}{c}{$V^k_{3}$} & \multicolumn{1}{c}{$V^k_{4}$} 
 & \multicolumn{1}{c}{$V^k_{5}$} & \multicolumn{1}{c}{$V^k_{6}$}
 & \multicolumn{1}{c}{$V^k_{7}$} \\
\hline
 ${^1S_0}$ & $\to$ & ${^1S_0}$ & $f$
 & 0       & -0.1419 & -0.0651 & -0.0089 & 0.0282 & -0.1631 & -0.1051 \\
           & $\to$ & ${^3P_0}$ & $g$
 & 0       &  0.0005 & -0.0093 & -0.0052 & 0.0083 & -0.0867 & -0.0543 \\
           &       &           & $h$
 & 0       &  0.0005 &  0.0035 &  0.0025 &-0.0212 &  0.1265 &  0.0800 \\
\hline
\hline
 \multicolumn{3}{c}{$\Lambda \Lambda \to \Sigma^- p$} & $k$ 
 & \multicolumn{1}{c}{$V^k_{1}$} & \multicolumn{1}{c}{$V^k_{2}$} 
 & \multicolumn{1}{c}{$V^k_{3}$} & \multicolumn{1}{c}{$V^k_{4}$} 
 & \multicolumn{1}{c}{$V^k_{5}$} & \multicolumn{1}{c}{$V^k_{6}$}
 & \multicolumn{1}{c}{$V^k_{7}$} \\
\hline
 ${^1S_0}$ & $\to$ & ${^1S_0}$ & $f$
 & 0       &  0.1004 &  0.0460 & -0.0009 & -0.0199 &  0.0720 &  0.1177 \\
           & $\to$ & ${^3P_0}$ & $g$
 & 0       &  0.0011 &  0.0065 &  0.0037 & -0.0034 &  0.0391 &  0.0606 \\
           &       &           & $h$
 & 0       &  0.0011 & -0.0024 &  0.0043 &  0.0144 & -0.0567 & -0.0894 \\
\hline
\hline
\end{tabular} 
\end{center}
\label{TAB:Coef}
\end{table}

\vspace*{2pt}
\section{Nonmesonic Decay of Double-$\Lambda$ Hypernuclei \label{SEC:Res}}
Now we show the results of $\Lambda \Lambda \to YN$ decay rates of ${^{\phantom{\Lambda}6}_{\Lambda \Lambda}}$He.
In Table \ref{TAB:He6LL},  
 the $\Lambda \Lambda \to YN$ transition is estimated by extending the $\pi+K+$DQ model to the $S=-2$ system.

For the case of $\Lambda \Lambda \rightarrow \Lambda n$ transition,
 the OPE process is forbidden 
 because of the isospin symmetry on the strong vertex.
Although the short-range interaction becomes dominant, 
 their contributions are drastically reduced by the short range correlation.
The DQ contribution is much larger than the kaon exchange contribution for the PV transition channel,
 but it is about $0.1\%$ of the free $\Lambda$ decay rate.
In our full calculation,
 we find that the DQ reduces the total PC transition rate to almost zero.
In contrast,
 the meson exchange and the DQ are added coherently in the PV channels, 
 resulting in a relatively large PV transition rate, which is about $0.2\%$ of the free $\Lambda$ decay.

The OPE contributions to the $\Lambda \Lambda \to \Sigma N$ transition
 is expected to be small in the $J=0$ transition.
In fact,
 some previous analyses have revealed that 
 the $J=0$ decay rate is only few $\%$ of the $J=1$ transition rate. 
Table~\ref{TAB:He6LL} shows that the decay rate in OPE is about $1\%$ of the free $\Lambda$ decay
 in $\Lambda \Lambda \to \Sigma^- p$ transition.
The kaon exchange process additively contributes to the OPE process, and the decay rate
 becomes about twice as large as the OPE alone in both the PC and PV channels.
Thus the $\Lambda \Lambda \rightarrow \Sigma N$ decay rate in meson exchange processes 
 becomes much larger than the $\Lambda \Lambda \rightarrow \Lambda n$.
The DQ transition rate for the $\Lambda \Lambda \rightarrow \Sigma N$ transition
 is also larger than the $\Lambda \Lambda \rightarrow \Lambda n$ case, and 
 it behaves differently from the $\Lambda \Lambda \rightarrow \Lambda n$ transition.
In our full calculation,
 the PV transition rate is largely reduced, while the PC is enhanced due to the DQ contribution.
Therefore
 the $\Lambda \Lambda \rightarrow \Sigma N$ transition occurs mainly in the PC channel.

It is also interesting to check the breaking of the empirical $\Delta I=1/2$ rule 
 for the $\Lambda \Lambda \to YN$ transition,
 which leads to
\begin{equation}
\Gamma_{\Sigma^- p} = 2 \Gamma_{\Sigma^0 n}.
\label{EQ:D12}
\end{equation}
This relation shows that the nonmesonic weak decay of double-$\Lambda$ hypernuclei provides us with
 an opportunity of direct confirmation of the validity of the empirical $\Delta I=1/2$ dominance.
In our prediction,
 this relation is badly violated due to the DQ contribution, in which 
 the $\Delta I=3/2$ amplitude exists through the weak four-quark Hamiltonian.
This situation is similar to the $\Lambda N \to NN$ transition.
Thus the test of the $\Delta I=1/2$ rule in the $\Lambda \Lambda \to YN$ transition
 may reveal the importance of the short-range part in the weak baryonic interaction,
 and may also help to understand the mechanism of the $\Delta I=1/2$ enhancement observed
 in the nonleptonic weak decays of the free hyperons and kaons.

\begin{table}[htb]
\caption{$\Lambda \Lambda \to YN$ decay rates of ${^{\hspace*{2.5mm} 6}_{\Lambda \Lambda}}$He
         in unit of $\Gamma_\Lambda$.
         The PC and PV denote the ${^1}S_0 \to {^1}S_0$ parity-violating and 
         ${^1}S_0 \to {^3}P_0$ parity-conserving transitions, respectively.} 
\begin{center}
\begin{tabular}{cccp{0.1mm}ccp{0.1mm}cc}
\hline \hline
 & \multicolumn{2}{c}{$\Gamma_{\Lambda n}$}  &
 & \multicolumn{2}{c}{$\Gamma_{\Sigma^0 n}$} &
 & \multicolumn{2}{c}{$\Gamma_{\Sigma^- p}$} \\
\cline{2-3} \cline{5-6} \cline{8-9}
\raisebox{4mm}[0pt][0pt]{${^{\phantom{\Lambda}6}_{\Lambda \Lambda}}$He} 
& PC & PV &&  PC & PV &&  PC & PV \\
\hline
$\pi$      & -----  & -----  && 0.0004 & 0.0030 && 0.0009 & 0.0061 \\
$\pi+K$    & 0.0001 & 0.0002 && 0.0011 & 0.0040 && 0.0021 & 0.0079 \\
DQ         & 0.0001 & 0.0012 && 0.0047 & 0.0037 && 0.0012 & 0.0018 \\
$\pi+K$+DQ & 0.0000 & 0.0024 && 0.0065 & 0.0000 && 0.0064 & 0.0021 \\
\hline
total      & \multicolumn{2}{c}{0.0024} && \multicolumn{2}{c}{0.0065}
           && \multicolumn{2}{c}{0.0085} \\
\hline
\hline
Ref.\cite{IUM:NPA}    
           & \multicolumn{2}{c}{0.025} && \multicolumn{2}{c}{0.0006}
           && \multicolumn{2}{c}{0.0012} \\
Ref.\cite{PRB:PRC}(E)
           & \multicolumn{2}{c}{0.0044} && \multicolumn{2}{c}{0.011}
           && \multicolumn{2}{c}{0.022} \\
Ref.\cite{PRB:PRC}(F)
           & \multicolumn{2}{c}{0.036} && \multicolumn{2}{c}{0.001}
           && \multicolumn{2}{c}{0.003} \\
\hline
\hline
\end{tabular}
\end{center}
\label{TAB:He6LL}
\end{table}

We compare our results with the previous calculations carried out
 by Itonaga et al. \cite{IUM:NPA} and Parre\~no et al. \cite{PRB:PRC}.
We all agree the bare pion-exchange amplitudes in the $\Lambda \Lambda \to \Sigma N$ transitions
 \cite{IP:Pdis}.
In Table~\ref{TAB:He6LL},
Ref.~\cite{IUM:NPA} includes effects of two-pion exchange in the scalar ($\sigma$) channel as well as
 the vector ($\rho$) channel.
Ref.~\cite{PRB:PRC} contains all the pseudoscalar mesons and the vector mesons exchanged between
 $\Lambda \Lambda$.
Ref.\cite{PRB:PRC} took two different approaches for the final state interaction.
The full calculation (F) is based on the scattering $T$ matrix in the final states, while
 an effective treatment (E) evaluates the transition amplitudes by multiplying 
 a correlation function to the relative wave function.
They pointed out that the full calculation enhances $\Gamma_{\Lambda n}$
 while it reduces $\Gamma_{\Sigma N}$.
The effective treatment does not include the conversion from $\Sigma N$ to $\Lambda N$.

In the present work, we make an effective treatment of the final state interactions. 
Therefore, our results have to be compared to the ones in Ref.\cite{PRB:PRC}(E). 
Our calculation is similar to their effective treatment and thus the $\Gamma_{\Lambda n}$ enhancement
 due to the final state interaction is not shown.
The total decay rates in our results are reduced by a factor about two due to the larger
 $b$ parameter for the $\Lambda$ wave function.
We note that if the same $b$ value is used, our results agree with the (E) results in Ref.\cite{PRB:PRC}.
We also point out that the ratio $\Gamma_{\Sigma^- p}/\Gamma_{\Sigma^0 n}$ is fixed to $2$
 by the $\Delta I=1/2$ rule in the meson exchange calculations.

\section{Conclusion \label{SEC:Con}}
In the double $\Lambda$ hypernuclear system,
 the novel decay mode, $\Lambda \Lambda \to YN$, is allowed.
The hyperon-induced decay rates are much smaller than the decay rate of free $\Lambda$
 ($\Gamma_\Lambda$).
This is because the $\Lambda \Lambda \to YN$ transition occurs only in the $J=0$ channel.
Although the $\Lambda \Lambda \to YN$ transition is not easy to detect
 experimentally, it is important to extract the information about the $J=0$ transition.

Our result shows that the DQ process gives significant contribution to 
 the $\Lambda \Lambda \to YN$ transition.
It strongly suppresses the PC amplitude for the $\Lambda \Lambda \to \Lambda n$ transition
 and the PV amplitude for the $\Lambda \Lambda \to \Sigma  N$ transition.
Furthermore 
 the DQ largely breaks the empirical $\Delta I=1/2$ rule.
Unlike the $\Lambda N \to NN$ transition,
 we have an opportunity to check directly the breaking of $\Delta I=1/2$ dominance
 for the $\Lambda \Lambda \to \Sigma  N$ transition.

\section*{Acknowledgment}
\noindent
The authors acknowledge Profs. A.~Parre\~no, K.~Itonaga, H.~Nemura, and Y.~Akaishi for variable discussions.
One of authors, K. S., thanks to JSPS Research Fellowship for financial support.
This work is supported in part by Grants-in-Aid for Scientific Research
 (No.11640261) from Japan Society for the Promotion of Science (JSPS).

\end{document}